\documentclass[fleqn,twoside,11pt]{article}

\usepackage{amssymb}
\usepackage{graphicx}

\usepackage{amsmath}

\begin{document}
\noindent
{\sf University of Shizuoka}

\hspace*{11cm} {\large US-05-07}\footnote{
To appear in Phys.~Rev. D. (2006)}\\

\vspace{-5mm}
\hspace*{11cm} {\large Revised Version of}\\

\vspace{-5mm}
\hspace*{11cm} {\large  US-05-01(hep-ph/0502054)} \\

\vspace{-5mm}
\vspace{2mm}

\begin{center}

{\Large\bf Shape of the Unitary Triangle}\\[.2in]
{\Large\bf and Phase Conventions of the CKM Matrix}

\vspace{3mm}

{\bf Yoshio Koide}

{\it Department of Physics, University of Shizuoka, 
52-1 Yada, Shizuoka 422-8526, Japan\\
E-mail address: koide@u-shizuoka-ken.ac.jp}

\date{\today}
\end{center}

\vspace{3mm}
\begin{abstract}
A shape of the unitary triangle versus a $CP$ violating parameter
$\delta$ depends on the phase conventions of the CKM matrix,
because the $CP$ violating parameter $\delta$ cannot directly 
be observed, so that it is not rephasing-invariant.
In order to seek for a clue to the quark mass matrix structure
and the origin of the $CP$ violation, the dependence of the
unitary triangle shape on the parameter $\delta$ is 
systematically investigated.
\end{abstract}



\vspace{5mm}


\noindent{\large\bf 1 \ Introduction} \ 

Usually, it is taken that any phase conventions of the 
Cabibbo-Kobayashi-Maskawa (CKM) \cite{Cabibbo,KM} matrix are 
equivalent to
each other because of the rephasing invariance.
This is true, as far as the observable quantities are concerned.
However, quark mass matrices $(M_u, M_d)$ are not rephasing
invariant, although
those are invariant under rebasing: $M_u \rightarrow
M'_u = A^\dagger M_u B_u$, $M_d \rightarrow
M'_d = A^\dagger M_d B_d$.
Sometimes, rephasing invariance is confused with rebasing
invariance.
Most experimentalists have an interest in relations among the observed
values (masses $m_{qi}$ and CKM parameters $|V_{ij}|$), 
which are rephasing invariant.
On the other hand, most model-builders take an interest 
in relations
between mass matrix parameters and observable quantities,
where those relations are model-dependent and are not
rephasing-invariant.
Usually, model-builders put some ansatz on the mass matrices
$(M_u, M_d)$,
which are given on a specific flavor basis.
Then, the ansatz will give a constraint on the $CP$ violating
phases of the CKM matrix $V=U_{uL}^\dagger U_{dL}$.
We would like to emphasize that a $CP$ violating parameter
$\delta$ in the CKM matrix is not observable, and it
depends on the phase convention of the CKM matrix
(so that it depends on a mass matrix model).
The observable quantities which are related to
$CP$ violation are angles $(\phi_1, \phi_2, \phi_3)
=(\beta, \alpha, \gamma)$ in the unitary triangle which are
defined in Eq.~(1.3) later.
Only when we take a specific phase convention, the 
parameter $\delta$ becomes observable, for example,
such as the $\delta_{13}$ parameter in the standard phase
convention \cite{SD-CKM} of the CKM matrix.
To investigate a phase convention with a reasonable value
of $\delta$ means to investigate a corresponding specific
flavor basis on which a quark mass matrix model is described,
although it is not directly.

For example, by noticing that predictions based on the maximal $CP$ 
violation hypothesis \cite{maxCP}
depend on the phase convention, the author \cite{Koide_maxCP} has 
recently pointed out that 
we can obtain successful predictions on the unitary triangle
only when we adopt the original Kobayashi-Maskawa (KM) \cite{Cabibbo} 
phase convention and the Fritzsch-Xing \cite{V33} phase convention.
If we put the ansatz on the standard phase convention 
\cite{SD-CKM} of the CKM matrix,
we will obtain wrong results on the unitary triangle.
For experimental studies, what convention we adopt is not
important, but, for model-building of the quark and lepton mass
matrices, it is a big concern.
In the present paper, in order to look for a clue to
the origin of the $CP$ violating phase $\delta$ 
 (what elements in the quark mass matrices contain the
$CP$ violating phase $\delta$ and how the magnitude of 
$\delta$ is), 
we will systematically investigate
whole phase conventions of the CKM matrix, comparing with
the present experimental data of the unitary triangle.

Recent remarkable progress of the experimental $B$ physics
\cite{B} has put the shape of the unitary
triangle within our reach. 
The world average value of the angle $\beta$ \cite{PDG04} which has been
obtained from $B_d$ decays is
$$
\sin 2\beta = 0.736 \pm 0.049 \ \ 
\left( \beta = 23.7^\circ {}^{+2.2^\circ}_{-2.0^\circ} \right) ,
\eqno(1.1)
$$
and the best fit \cite{PDG04} for the CKM matrix  $V$ also gives 
$$
\gamma = 60^{\circ} \pm 14^{\circ} \ , \ \ \ 
\beta = 23.4^{\circ} \pm 2^{\circ} \ ,
\eqno(1.2)
$$
where the angles $\alpha$, $\beta$ and $\gamma$ are defined by
$$
\alpha \equiv \phi_2 = {\rm Arg} 
\left[-\frac{V_{31} V^{*}_{33}}{V_{11} V^{*}_{13}} \right]
 , \ \ \ 
\beta \equiv \phi_1 = {\rm Arg} 
\left[-\frac{V_{21} V^{*}_{23}}{V_{31} V^{*}_{33}} \right]
 , \ \ 
\gamma \equiv \phi_3 = {\rm Arg} 
\left[-\frac{V_{11} V^{*}_{13}}{V_{21} V^{*}_{23}} \right]
 .
\eqno(1.3)
$$
Also we know the observed values \cite{PDG04} of the magnitudes 
$|V_{ij}|$ of the CKM matrix elements:
$$
|V_{us}| = 0.2200 \pm 0.0026 , \ \
|V_{cb}| = 0.0413 \pm 0.0015 , \ \
|V_{ub}| = 0.00367 \pm 0.00047 ,
\eqno(1.4)
$$
$$
{\rm Re} V_{td} = 0.0067 \pm 0.0008 , \ \ \ 
{\rm Im} V_{td} = -0.0031 \pm 0.0004 .
\eqno(1.5)
$$
Thus, nowadays, we have almost known the shape of the unitary triangle
$V^{*}_{ud} V_{ub} + V^{*}_{cd} V_{cb} + V^{*}_{td} V_{tb} =0$.
We are interested 
what logic can give the observed magnitude of the $CP$ violation. 

There are, in general, 9 independent phase conventions \cite{FX-9V}
of the CKM matrix.
In the present paper, we define the expressions of the CKM matrix $V$ as
$$
V = V(i,k) \equiv R^T_i P_j R_j R_k \ \ \ \ \ (i \neq j \neq k) ,
\eqno(1.6)
$$
where 
$$
R_1 (\theta) = \left(
\begin{array}{ccc}
1 & 0 & 0 \\
0 & c & s \\
0 & -s & c 
\end{array} \right) , \ \ \ \ 
R_2 (\theta) = \left(
\begin{array}{ccc}
c & 0 & s \\
0 & 1 & 0 \\
-s & 0 & c 
\end{array} \right) , \ \ \ \ 
R_3 (\theta) = \left(
\begin{array}{ccc}
c & s & 0 \\
-s & c & 0 \\
0 & 0 & 1 
\end{array} \right) ,
\eqno(1.7)
$$
($s=\sin\theta$ and $c=\cos\theta$)
and 
$$
P_1 = {\rm diag} (e^{i \delta}, \ 1, \ 1) , \ \ 
P_2 = {\rm diag} (1, \ e^{i \delta}, \ 1), \ \ 
P_3 = {\rm diag} (1, \ 1, \ e^{i \delta}) .
\eqno(1.8)
$$
The expressions $V(1,3)$, $V(1,1)$ and $V(3,3)$ 
correspond to the standard \cite{SD-CKM}, original KM 
\cite{KM} and Fritzsch-Xing \cite{V33}
phase conventions, respectively.

By the way, the CKM matrix structure (1.6) is related to 
a quark mass matrix model under the following specific assumption:
We assume that the phase factors in the quark mass matrices 
$M_f$ \ $(f=u,d)$ can be 
factorized by the phase matrices $P_f$ as
$$
M_f = P^{\dagger}_{fL} \widetilde{M}_f P_{fR} \ ,
\eqno(1.9)
$$
where $P_f$ are phase matrices and $\widetilde{M}_f$ are real matrices. 
(This is possible  for a mass matrix which has specific zero-textures,
for example, such as a model with  nearest-neighbor interactions (NNI)
\cite{NNI}. For details, see Appendix.) 
The real matrices $\widetilde{M}_f$ are diagonalized by rotation (orthogonal)
matrices $R_f$ as
$$
R^{\dagger}_f \widetilde{M}_f R_f = D_f 
\equiv {\rm diag} (m_{f1}, \ m_{f2}, \ m_{f3} ),
\eqno(1.10)
$$
[for simplicity, we have assumed that $M_f$ are Hermitian (or symmetric) 
matrix, i.e. $P_{fR} = P_{fL}$ (or $P_{fR} = P_{fL}^{\ast}$)], so that 
the CKM matrix $V$ is given by
$$
V = R^T_u P R_d \ ,
\eqno(1.11)
$$
where $P = P^{\dagger}_{uL} P_{dL}$. 
The quark masses $m_{fi}$ are only 
determined by $\widetilde{M}_f$. 
In other words, the rotation parameters are
given only in terms of the quark mass ratios, and independent 
of the $CP$ violating phases.
In such a scenario, the $CP$ violation parameter 
$\delta$ can be adjusted  without changing the quark mass values.
In the present paper,  by fixing the rotation matrices $R_u$ and $R_d$ 
(i.e. by fixing the quark masses), we tacitly assume that the 
$CP$ violation is described only by the adjustable parameter $\delta$.
Then, the expression of the law of the $CP$ violation depends on the 
phase conventions of the CKM matrix.

For example, the phase convention $V(2,3)$ 
$$
V(2,3) = R_2^T (\theta_{13}^u) P_1 (\delta) R_1 (\theta_{23}) 
R_3(\theta^d_{12}),
\eqno(1.12)
$$
suggests the quark mass matrix structures
$$
\begin{array}{c}
\widetilde{M}_u = R_1(\theta^u_{23}) R_2(\theta_{13}^u) 
D_u R^T_2(\theta_{13}^u)
R_1^T (\theta^u_{23}) \ , \\
\widetilde{M}_d = R_1(\theta^d_{23}) R_3(\theta_{12}^d) 
D_d R^T_3(\theta_{12}^d)
R_1^T (\theta^d_{23}) \ ,
\end{array}
\eqno(1.13)
$$
with $\theta_{23}=\theta^d_{23}-\theta^u_{23}$. 
Therefore, in order to seek for a clue to the quark mass matrix
structure, we interest in the relations of the phase conventions
(1.6) to the observed unitary triangle shape.

\vspace{3mm}

\noindent{\large\bf 2  \ Rephasing invariant quantity $J$ versus $\delta$} 

Of the three unitary triangles $\bigtriangleup^{(ij)}$
[$(ij)=(12)$, $(23)$, $(31)$] which denote the unitary conditions
$$
\sum_k V^{*}_{ki} V_{kj} = \delta_{ij}  ,
\eqno(2.1)
$$
we usually discuss the triangle $\bigtriangleup^{(31)}$, i.e.
$$
V^{*}_{ud} V_{ub} + V^{*}_{cd} V_{cb} + V^{*}_{td} V_{tb} =0,
\eqno(2.2)
$$
because the triangle $\bigtriangleup^{(31)}$ is the most useful 
one for the experimental studies.

The rephasing invariant quantity \cite{J} $J$  is given by
$$
J=\frac{|V_{i1}||V_{i2}||V_{i3}||V_{1k}||V_{2k}||V_{3k}|}
{(1 -|V_{ik}|^2 ) |V_{ik}|} \sin \delta  ,
\eqno(2.3)
$$
in the phase convention $V(i,k)$, 
where the $CP$ violating phase $\delta$ has been defined by Eq.~(1.7).
(We again would like to emphasize that the parameter $\delta$ is not
observable in the direct meaning, and it is model-dependent. 
As we stated in Sec.~1, the observable quantities which are related to
$CP$ violation are angles $(\phi_1, \phi_2, \phi_3)
=(\beta, \alpha, \gamma)$ in the unitary triangle.)
Note that the 5 quantities (not 6 quantities) $|V_{i1}|$, $|V_{i2}|$, 
$|V_{i3}|$, $|V_{1k}|$, $|V_{2k}|$ and $|V_{3k}|$ in the expression
$V(i,k)$ are independent of the phase parameter $\delta$.
(In other words, only the remaining 4 quantities are dependent 
of $\delta$.)
Therefore, the rephasing invariant quantity $J$ is dependent 
on the parameter $\delta$ only through the factor $\sin\delta$.
A ``maximal $CP$ violation" means a maximal $J$, so that it
means a maximal $\sin\delta$.
Thus, the maximal $CP$ violation hypothesis depends on the
phase conventions.

{}From the expression (2.3), for the observed fact 
$1 \gg |V_{us}|^2 \simeq |V_{cd}|^2 \gg |V_{cb}|^2\simeq 
|V_{ts}|^2 \gg |V_{ub}|^2$, the rephasing invariant quantity
$J$ is classified in the following four types:
$$
\begin{array}{ll}
(A): & J \simeq |V_{ub}||V_{td}| \sin \delta  , \\
(B): & J \simeq |V_{us}||V_{cb}||V_{ub}| \sin \delta  ,\\
(C): & J \simeq |V_{us}||V_{cb}||V_{td}| \sin \delta  ,\\
(D): & J \simeq |V_{cb}|^2 \sin \delta  .
\end{array}
\eqno(2.4)
$$
The corresponding phase conventions $V(i,k)$ are listed in
Table 1.

The present experimental values (1.2) suggest 
$\alpha \simeq 90^\circ$.
Since only the cases $V(1,1)$ and $V(3,3)$ can give 
$\delta \simeq \alpha$ as seen in Table 1, 
the ``maximal $CP$ violation hypothesis" (i.e. maximal
$\sin\delta$ hypothesis) can give successful results
only for the cases $V(1,1)$ and $(3,3)$ \cite{Koide_maxCP}.

\vspace{3mm}

\noindent{\large\bf 3 \ Angles $\phi_i$ versus $\delta$} 

In the present section, we systematically investigate
the relations between the angles $\phi_\ell$ ($\ell=1,2,3$)
and the $CP$ violating phase $\delta$ for each case
$V(i,k)$.

The angles $(\phi_1, \phi_2, \phi_3)\equiv (\beta, \alpha, \gamma)$ 
on the unitary triangle $\bigtriangleup^{(31)}$ are given 
by the sine rule
$$
\frac{r_1}{\sin\phi_1}=\frac{r_2}{\sin\phi_2}=
\frac{r_3}{\sin\phi_3}= 2 R,
\eqno(3.1)
$$
where $R$ is the radius of the circumscribed circle of the
triangle $\bigtriangleup^{(31)}$, and $r_i$ are defined by
$$
r_1 =|V_{13}| |V_{11}|, \ \ r_2 =|V_{23}| |V_{21}|, \ \ 
r_3 =|V_{33}| |V_{31}|.
\eqno(3.2)
$$
Then, the quantity $J$ is rewritten as follows:
$$
J = 2 r_m r_n \sin\phi_\ell
 = \frac{1}{R} r_\ell r_m r_n = 
\frac{1}{R} |V_{11}| |V_{21}| |V_{31}| |V_{13}| |V_{23}| |V_{33}| ,
\eqno(3.3)
$$
where $(\ell, m, n)$ is a cyclic permutation of (1,2,3).
{}From Eqs.~(2.3), (3.1) and (3.3), the angles $\phi_\ell$ 
are given by the formula
$$
\sin\phi_\ell=\frac{|V_{i1}||V_{i2}||V_{i3}|
|V_{1k}||V_{2k}||V_{3k}|\sin\delta}{|V_{m1}||V_{m3}|
|V_{n1}||V_{n3}|(1 -|V_{ik}|^2 ) |V_{ik}|} .
\eqno(3.4)
$$

Of the three sides in the expression $V(i,k)$, only one side $r_i$ 
is always independent of the phase parameter $\delta$.
And, of the three angle $\phi_i$, only one (we express it with
$\phi_\ell$), except for the
case $V(2,2)$, is approximately equal to the phase parameter 
$\delta$. In Table 1, we also list the side $r_i$ which is
independent of $\delta$ and the angle $\phi_\ell$ which is approximately
equal to $\delta$.

The relations between $\phi_i$  ($i=1,2,3$) and $\delta$ are 
illustrated in Figs.~1--8.
The curves have been evaluated by using the explicit expression
(1.6) (not by using the formula (3.4)).
In general, there are five $|V_{ij}|$ which are independent of
the phase parameter $\delta$.
For the cases that $|V_{us}|$, $|V_{cb}|$ and $|V_{ub}|$ are
$\delta$-independent $V_{ij}$, we have used the observed values
(1.4) as the input values, i.e. $|V_{us}|=0.22$, $|V_{cb}|=0.0413$
and $|V_{ub}|=0.00367$.
When $|V_{us}|$ ($|V_{cb}|$) is $\delta$-dependent,
but $|V_{cd}|$ ($|V_{ts}|$) is $\delta$-independent,
we have, for convenience,  used the input values $|V_{cd}|=0.22$ 
($|V_{ts}|=0.0413$).
When $|V_{ub}|$ is $\delta$-dependent,
but $|V_{td}|$ is $\delta$-independent,
we have, for convenience, used the input values $|V_{td}|=0.0084$,
which is a predicted value of $|V_{td}|$ 
in the case $V(1,1)$ with the maximal $\sin\delta$. 
However, for the case $V(2,2)$, since both $|V_{ub}|$ and
$|V_{td}|$ are $\delta$-dependent, so that we cannot use such an 
approximate substitute.
As seen in Table 1, the case $V(2,2)$ needs a small value of 
$\delta$ compared with other cases, so that the case is not so
interesting.
We omit the case $V(2,2)$ from the present study.

As seen in Figs.~1--8, of the maximal values of the
three $\sin\phi_i$ ($i=1,2,3$), two can take 
$(\sin\phi_i)_{max}=1$, while one  (we express it with 
$\phi_s$) always takes a smaller value 
than one, i.e. $(\sin\phi_s)_{max} < 1$.
The angle $\phi_s$ with  $(\sin\phi_s)_{max} < 1$
is $\phi_1$ for the cases A and B, and is $\phi_3$ for 
the case C.
If we assume that nature chooses the value of the phase 
parameter $\delta$ such as $\sin\phi_s$ is maximal,
as shown in Table 2,
the cases $V(i,k)$ with $i \neq k$ can predict reasonable values 
of the angles $\phi_i$ ($i=1,2,3$).

More straightforward ansatz is as follow:
the value of 
$\sin\alpha$ has to take its maximal value $\sin\alpha=1$.
Then, all cases $V(i,k)$ can give reasonable values of the angles
as seen in Table 2.
However, this ansatz is merely other expression of the
observed fact (1.2).
In the maximal $CP$ violation hypothesis, the hypothesis
has been imposed on the $CP$ violating phase parameter
$\delta$, which is not a directly observable quantity.
Therefore, the hypothesis could choose specific phase
conventions $V(1,1)$ and $V(3,3)$ (consequently, 
specific quark mass matrix structures) as experimentally
favorable ones.
In contrast to the maximal $CP$ violation hypothesis,
the ansatz for the directly observable quantities such
as $(\sin\alpha)_{max}=1$ cannot choose a specific phase
convention $V(i,k)$ as a favorable one.
It is unlikely that the ansatz $\sin\alpha=1$ gives a clue to
the origin of the $CP$ violating phase in the quark mass
matrices.

\vspace{3mm}

\noindent{\large\bf 4 \ Radius of the circumscribed circle}

When we see the unitary triangle from the geometrical point of view, 
we find that the triangle 
$\bigtriangleup^{(31)}$ has the plumpest shape compared with
other triangles $\bigtriangleup^{(12)}$ and $\bigtriangleup^{(23)}$, 
so that the triangle $\bigtriangleup^{(31)}$  has the shortest radius 
$R_{min}$ of the circumscribed circle compared with 
the other cases $\bigtriangleup^{(12)}$ and $\bigtriangleup^{(23)}$.
Therefore, let us put the following assumption:
the phase parameter $\delta$ takes the value so that the radius of the
circumscribed circle $R(\delta)$ takes its minimum value.
The radius $R(\delta)$  is given by the sine rule (3.1).
Note that the side $r_i$ in the expression $V(i,k)$ is independent of
the parameter $\delta$. 
Therefore, the minimum of the radius $R(\delta)$ means 
the maximum of $\sin\phi_i(\delta)$ in the phase convention $V(i,k)$.
In Table 3, we list values of $(\phi_1, \phi_2, \phi_3)$ at 
$\delta=\delta_0$
at which $\sin\phi_i$ takes its maximal value.
As seen in Table 3, all cases except for $V(1,1)$ and $V(3,3)$ (and also
$V(2,2)$) can give favorable predictions.
Therefore, this ansatz is also not useful to select a specific $V(i,k)$.

If we put further strong constraint that 
the phase parameter $\delta$ takes own value so that $\sin\phi_i(\delta)$
takes its maximal value $\sin\phi_i=1$, then, we find that 
the possible candidates are only two: $V(2,3)$ and $V(2,1)$.
(The other cases cannot take the value $\sin\phi_i=1$ under the observed
values (1.4) of $|V_{us}|$, $|V_{cb}|$ and $|V_{ub}|$.)
When we take account of the forms of the quark mass matrices 
$(M_u, M_d)$ which are 
suggested by Eq.~(1.11) from a specific phase convention $V(i,k)$,
we especially interest in the phase convention $V(2,3)$.
The phase convention (1.12) suggests the quark mass matrix
structure (1.13).
It is well known that if we require the zero-texture
$(M_d)_{11}=0$ for the down-quark mass matrix $M_d$,
we can obtain the successful prediction for $|V_{us}|$ \cite{Vus}
$$
|V_{us}|\simeq \sqrt{\frac{m_d}{m_s}} = 0.22.
\eqno(4.1)
$$
{}From the point of view of $M_u$-$M_d$ correspondence, 
if we also apply the zero-texture hypothesis to
the up-quark mass matrix $M_u$, we obtain
$$
|V_{ub}| \simeq s_{13}^u \simeq \sqrt{\frac{m_u}{m_t}} =0.0036,
\eqno(4.2)
$$
from $(M_u)_{11}=(m_{u3} - m_{u1}) c^u_{13} s^u_{13} c^u_{23}$, 
where we have used the quark mass values \cite{q-mass}
at $\mu=m_Z$.
The prediction is in excellent agreement with the observed
value (1.4).
(If we put $(M_u)_{11}=0$ on the mass matrix $M_u$ which is
suggested from the phase convention $V(3,3)$, we will obtain
$|V_{ub}/V_{cb}| \simeq \sqrt{m_u/m_c} = 0.059$, which is
in poor agreement with the observed value
$|V_{ub}/V_{cb}| =0.089^{+0.015}_{-0.014}$.)
Therefore, from the phenomenological point of view,
we are interested in the phase convention $V(2,3)$ rather
than the phase convention $V(3,3)$.

\vspace{3mm}

\noindent{\large\bf 5 \ Concluding remarks} 

In conclusion, we have investigated the dependence of the
unitary triangle shape on the $CP$ violating parameter
$\delta$ which is dependent on the phase conventions of
the CKM matrix.
The phase conventions are, generally, classified into the 9
expressions $V(i,k)$, Eq.~(1.6), which suggests
the quark mass matrix structures (1.9) with Eq.~(1.11).
If we require that the angle $\alpha$ ($\equiv\phi_2$) takes
$\sin\alpha=1$, all cases can predict favorable values of
$(\phi_1, \phi_2, \phi_3)$ as seen in Table 2.

However, we want to select a specific phase convention
$V(i,k)$ in order to seek for a clue to the quark mass
matrix structure and the origin of the $CP$ violation.
Then, the most naive and simplest hypothesis is the well-known
``maximal $CP$ violation hypothesis", which means
the requirement $\sin\delta=1$.
The ansatz selects the cases $V(1,1)$ and $V(3,3)$.
The relations between $V(3,3)$ and the quark mass matrices
$(M_u, M_d)$ have already discussed in Refs.~\cite{V33,Xing03}.

Another selection rule is a minimal circumscribed
circle hypothesis, which requires a maximal value of
$\sin\phi_i$ in the phase convention $V(i,k)$.
The hypothesis selects all cases except
for $V(i,i)$ ($i=1,2,3$) as favorable ones.
Only when we put a stronger constraint $\sin\phi_i=1$, we can
selects cases $V(2,3)$ and $V(2,1)$.
(In other cases, $\sin\phi_i$ cannot take $\sin\phi_i=1$
under the observed values (1.4) of $|V_{us}|$, $|V_{cb}|$ and
$|V_{ub}|$.)
We are interested in the case $V(2,3)$ because
the suggested quark mass matrices predict successful
relations $|V_{ub}|\simeq \sqrt{m_u/m_t}$ and 
$|V_{us}| \simeq \sqrt{m_d/m_s}$ under the simple
texture-zero hypotheses $(M_u)_{11}=0$ and 
$(M_d)_{11}=0$, respectively.

Although, in the present paper, we did not discuss
the neutrino mixing matrix \cite{MNS} $U=U^\dagger_{eL}U_{\nu L}$,
where $U_{eL}^\dagger M_e U_{eR}=D_e$ and 
$U_{\nu L}^\dagger M_\nu U^*_{\nu L} =D_\nu$,
the expressions $V(i,k)$ will also be useful for
studies of the neutrino mixings.
If we obtain data of $CP$ violation in the lepton sector
in the near future,
we can select a favorable expression $V(i,k)$ for the
mixing matrix $U$, and thereby, we will be able to get
a clue for investigating structures of $M_e$ and $M_\nu$
individually.

\newpage
\begin{center}
{\Large\bf Appendix:}\\
{\Large\bf Conditions on a mass matrix which is factorized}\\
{\Large\bf into a real matrix by phase matrices} 
\end{center}

We show that a mass matrix $M$ with a specific texture-zero
can always be factorized by phase matrices $P_L$ and $P_R$ as
$$
M = P_L^\dagger \widetilde{M} P_R ,
\eqno(A.1)
$$
where $\widetilde{M}$ is a real matrix, and
$$
P_L={\rm diag}(e^{i\delta_1^L}, e^{i\delta_2^L}, e^{i\delta_3^L}),
\ \ 
P_R={\rm diag}(e^{i\delta_1^R}, e^{i\delta_2^R}, e^{i\delta_3^R}).
\eqno(A.2)
$$
When we denote
$$
M_{ij} = |M_{ij}| e^{i \phi_{ij}} ,
\eqno(A.3)
$$
we obtain 9 relations
$$
\phi_{ij} = - (\delta_i^L - \delta_j^R).
\eqno(A.4)
$$
Although we have 6 parameters $\delta_i^L$ and  $\delta_i^R$,
the substantial number of the parameters is 5.
Therefore, we have 4 independent relations among the
phases $\phi_{ij}$.
In order that the phase parameters $\phi_{ij}$ are free each
other, 5 of 9 mass matrix elements must be zero.

Let us it in the concrete.
{}From the relations (A.4), we obtain
$$
\delta_1^L = \delta_1^R - \phi_{11} = \delta_2^R -\phi_{12}
=\delta_3^R -\phi_{13} ,
\eqno(A.5)
$$
$$
\delta_2^L = \delta_1^R - \phi_{21} = \delta_2^R -\phi_{22}
=\delta_3^R -\phi_{23} ,
\eqno(A.6)
$$
$$
\delta_3^L = \delta_1^R - \phi_{31} = \delta_2^R -\phi_{32}
=\delta_3^R -\phi_{33} .
\eqno(A.7)
$$
By eliminating $\delta_i^R$ from the relations (A.5) -- (A.7),
we obtain the following 4 independent relations among $\phi_{ij}$:
$$
\phi_{11}+\phi_{22}=\phi_{12}+\phi_{21} ,
\eqno(A.8)
$$
$$
\phi_{22}+\phi_{33}=\phi_{23}+\phi_{32} ,
\eqno(A.9)
$$
$$
\phi_{33}+\phi_{11}=\phi_{31}+\phi_{13} ,
\eqno(A.10)
$$
$$
\phi_{12}+\phi_{23} +\phi_{31} = 
\phi_{21}+\phi_{32} +\phi_{13}.
\eqno(A.11)
$$
If a matrix element $M_{ij}$ is zero, the corresponding
phase parameter $\phi_{ij}$ becomes unsettled.
Every relations (A.8) -- (A.11) contain such
unsettled phases more than one in order that the mass 
matrix $M$ can always be transformed into the real matrix 
$\widetilde{M}$ by phase matrices $P_L$ and $P_R$ as
Eq.~(A.1).
Therefore, 4 zero-textures are, at least, required.

Of course, if the phase parameters $\phi_{ij}$ satisfy
the relations (A.8) -- (A.11), the mass matrix $M$
can always be transformed into a real matrix $\widetilde{M}$
as Eq.~(A.1) without texture-zeros.

As such a typical mass matrix form which can be factorized
as Eq.~(A.1), a model with a NNI form \cite{NNI} is 
well-known:
$$
M = \left(
\begin{array}{ccc}
0 & a & 0 \\
a' & 0 & b \\
0 & b' & c \\
\end{array} \right) ,
\eqno(A.12)
$$
We should recall that
Branco, Lavoura and Mota \cite{Branco-Lavoura-Mota}
have shown that any quark mass
matrix form $(M_u, M_d)$ can be transformed into the NNI form
(A.12) by rebasing without losing generality.
However, even the mass matrix form $M_f$ in Eq.~(1.9) has 
a NNI form, in the present investigation, it means a case 
that the NNI form is an original form without rebasing.


\newpage

\begin{quotation}
{\bf Table 1 \ Classification of $V(i,k)$.}
The cases are classified under the approximation of
$1 \gg |V_{us}|^2 \simeq |V_{cd}|^2 \gg |V_{cb}|^2\simeq 
|V_{ts}|^2 \gg |V_{ub}|^2$. 
For the types of $J$, see Eq.~(2.8) in the text.
\end{quotation}

\begin{tabular}{|l|c|c|c|}\hline
Phase convention & Type of $J$ & $\delta$-independent $r_i$ & 
$\delta \simeq \phi_\ell$ \\ \hline
$V(1,1) =R_1^T P_2 R_2 R_1$ & A & $r_1$ & $\delta \simeq \phi_2$ \\
$V(3,3) =R_3^T P_1 R_1 R_3$ & A & $r_3$ & $\delta \simeq \phi_2$ \\ \hline
$V(1,2) =R_1^T P_3 R_3 R_2$ & B & $r_1$ & $\delta \simeq \phi_3$ \\
$V(1,3) =R_1^T P_2 R_2 R_3$ & B & $r_1$ & $\delta \simeq \phi_3$ \\
$V(2,3) =R_2^T P_1 R_1 R_3$ & B & $r_2$ & $\delta \simeq \phi_3$ \\ \hline
$V(2,1) =R_2^T P_3 R_3 R_1$ & C & $r_2$ & $\delta \simeq \phi_1$ \\
$V(3,1) =R_3^T P_2 R_2 R_1$ & C & $r_3$ & $\delta \simeq \phi_1$ \\
$V(3,2) =R_3^T P_1 R_1 R_2$ & C & $r_3$ & $\delta \simeq \phi_1$ \\ \hline
$V(2,2) =R_2^T P_1 R_1 R_2$ & D & $r_2$ & No simple relation \\ \hline
\end{tabular}

\vspace{2cm}

\begin{quotation}
{\bf Table 2 \ Maximal $\sin\phi_s$ hypothesis.}
\end{quotation}

\begin{tabular}{|c|c|ccccc|ccc|}\hline
  &       & \multicolumn{5}{c|}{$(\sin\phi_s)_{max}$  ($<1$)
at $\delta=\delta_0$} & \multicolumn{3}{c|}{$(\sin\phi_2)_{max}=1$ 
at $\delta=\delta_0$} \\
Type & $V(i,k)$ & $s$ & $\phi_1$ & $\phi_2 $ & $\phi_3$ & $\delta_0$ 
&  $\phi_1$ &  $\phi_3$ & $\delta_0$ \\ \hline
A & $V(1,1) $ & $s=1$ & $25.4^\circ$ & $64.6^\circ$ & 
$90.0^\circ$ & $115.3^\circ$ 
& $23.2^\circ$ & $66.8^\circ$ & $90.0^\circ$ \\
A & $V(3,3) $ &  $s=1$ & $23.2^\circ$ & $65.7^\circ$ & 
$91.1^\circ$ & $66.8^\circ$ 
& $21.4^\circ$ & $68.6^\circ$  & $91.1^\circ$ \\ \hline
B & $V(1,2)$ &  $s=1$ & $22.8^\circ$ & $91.0^\circ$ & 
$66.2^\circ$ & $114.8^\circ$
& $22.8^\circ$ & $67.2^\circ$ & $113.8^\circ$ \\
B & $V(1,3) $ &  $s=1$ & $23.2^\circ$ & $90.0^\circ$ & 
$66.8^\circ$ & $66.9^\circ$
&  $23.2^\circ$ &  $66.8^\circ$ & $66.9^\circ$ \\
B & $V(2,3)$ &  $s=1$ & $23.2^\circ$ & $90.0^\circ$ & 
$66.8^\circ$ & $113.2^\circ$
& $23.2^\circ$ &  $66.8^\circ$ & $113.2^\circ$ \\ \hline
C & $V(2,1) $ &   $s=3$ & $22.5^\circ$ & $90.0^\circ$ & 
$67.5^\circ$ & $157.5^\circ$
&  $22.5^\circ$ &  $67.5^\circ$ & $157.5^\circ$ \\
C & $V(3,1) $ &  $s=3$ & $25.7^\circ$ & $88.9^\circ$ & 
$65.4^\circ$ & $26.9^\circ$
& $24.6^\circ$ & $65.4^\circ$ & $25.7^\circ$ \\
C &$V(3,2) $ &  $s=3$ & $25.6^\circ$ & $88.9^\circ$ & 
$65.5^\circ$ & $153.3^\circ$
& $24.5^\circ$ &  $65.5^\circ$ & $154.4^\circ$  \\ \hline
\end{tabular}

\newpage
\vspace{2cm}

\begin{quotation}
{\bf Table 3 \ Minimal circumscribed circle hypothesis.}
The hypothesis requires a maximal
$\sin\phi_i$ in the phase convention $V(i,k)$.
The underlined values are obtained by the maximal $\sin\phi_i$
requirement.
\end{quotation}

\begin{tabular}{|c|c|cccc|}\hline
Type & $V(i,k)$ & $\phi_1$ & $\phi_2 $ & $\phi_3$ & $\delta_0$ 
 \\ \hline
A & $V(1,1) $ & \underline{$25.4^\circ$} & $64.6^\circ$ & 
$90.0^\circ$ & $115.3^\circ$ 
 \\
A & $V(3,3) $ & $23.2^\circ$ & $66.8^\circ$ &  
\underline{$90.0^\circ$} & $67.8^\circ$ 
 \\ \hline
B & $V(1,2)$ & \underline{$22.8^\circ$} & $91.0^\circ$ & 
$66.2^\circ$ & $114.8^\circ$
 \\
B & $V(1,3) $ & \underline{$23.2^\circ$} & $90.0^\circ$ & 
$66.8^\circ$ & $66.9^\circ$
 \\

B & $V(2,3)$ & $23.2^\circ$ & \underline{$90.0^\circ$} & 
$66.8^\circ$ & $113.2^\circ$
 \\ \hline
C & $V(2,1) $ &  $22.5^\circ$ & \underline{$90.0^\circ$} & 
$67.5^\circ$ & $157.5^\circ$
 \\
C & $V(3,1) $ & $25.7^\circ$ & $88.9^\circ$ & \underline{$65.4^\circ$} 
& $26.9^\circ$
 \\
C &$V(3,2) $ & $25.6^\circ$ & $88.9^\circ$ & \underline{$65.5^\circ$} & 
$153.3^\circ$
  \\ \hline
\end{tabular}

\newpage

\begin{figure}
\begin{center}

\includegraphics[width=8.6cm]{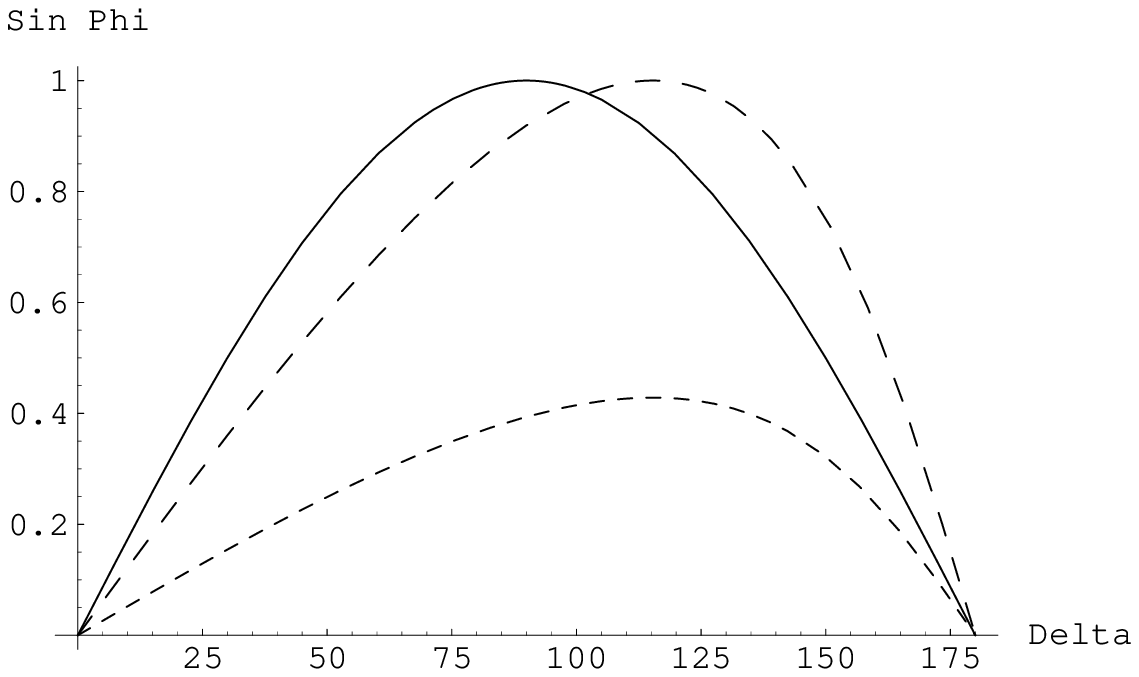}

\end{center}
\caption{
$\sin\phi_i$ ($i=1,2,3$) versus $\delta$ in $V(1,1)$.  
The curves $\sin\alpha$, $\sin\beta$ and $\sin\gamma$ are
denoted by a solid line, a dotted line  and  a dashed line,
respectively.
}
\label{v11}
\end{figure}

\begin{figure}
\begin{center}

\includegraphics[width=8.6cm]{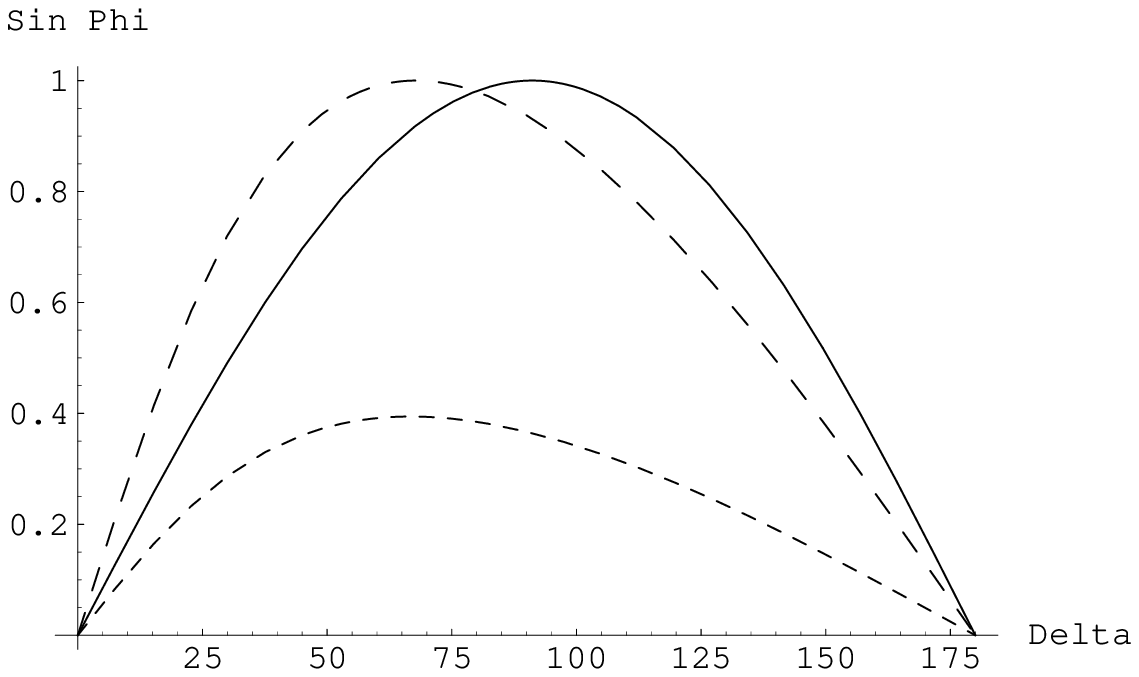}

\end{center}
\caption{
$\sin\phi_i$ ($i=1,2,3$) versus $\delta$ in $V(3,3)$.  
The curves $\sin\alpha$, $\sin\beta$ and $\sin\gamma$ are
denoted by a solid line, a dotted line  and  a dashed line,
respectively.
}
\label{v33}
\end{figure}

\begin{figure}
\begin{center}

\includegraphics[width=8.6cm]{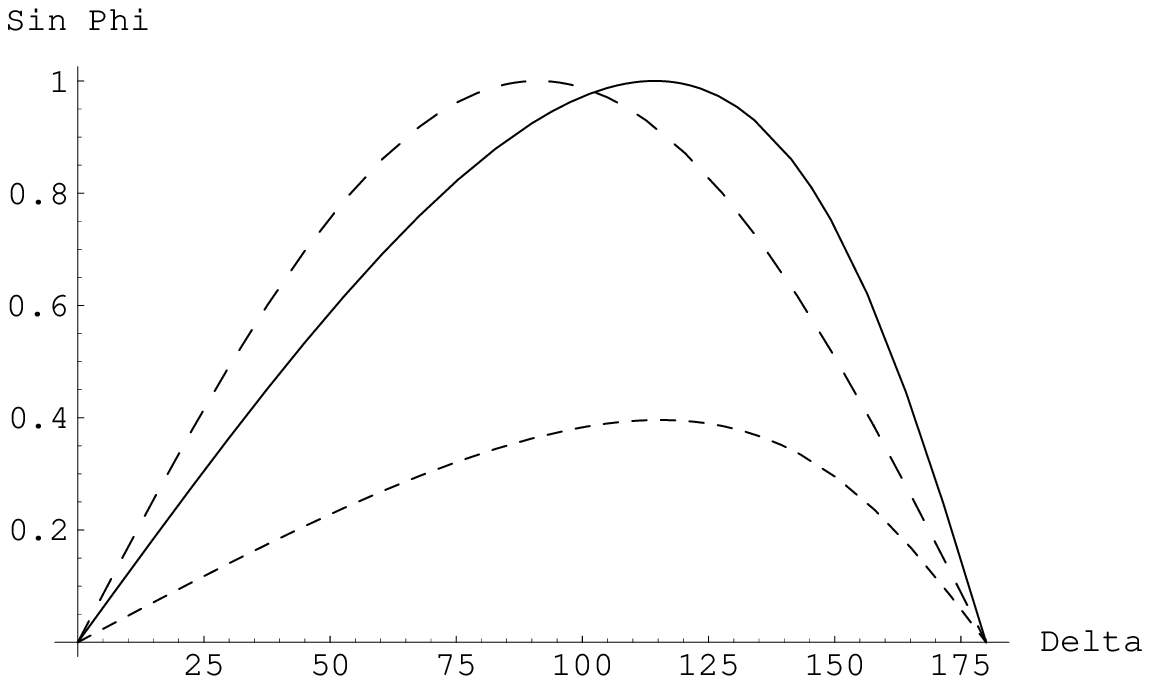}

\end{center}
\caption{
$\sin\phi_i$ ($i=1,2,3$) versus $\delta$ in $V(1,2)$.  
The curves $\sin\alpha$, $\sin\beta$ and $\sin\gamma$ are
denoted by a solid line, a dotted line  and  a dashed line,
respectively.
}
\label{v12}
\end{figure}

\begin{figure}
\begin{center}

\includegraphics[width=8.6cm]{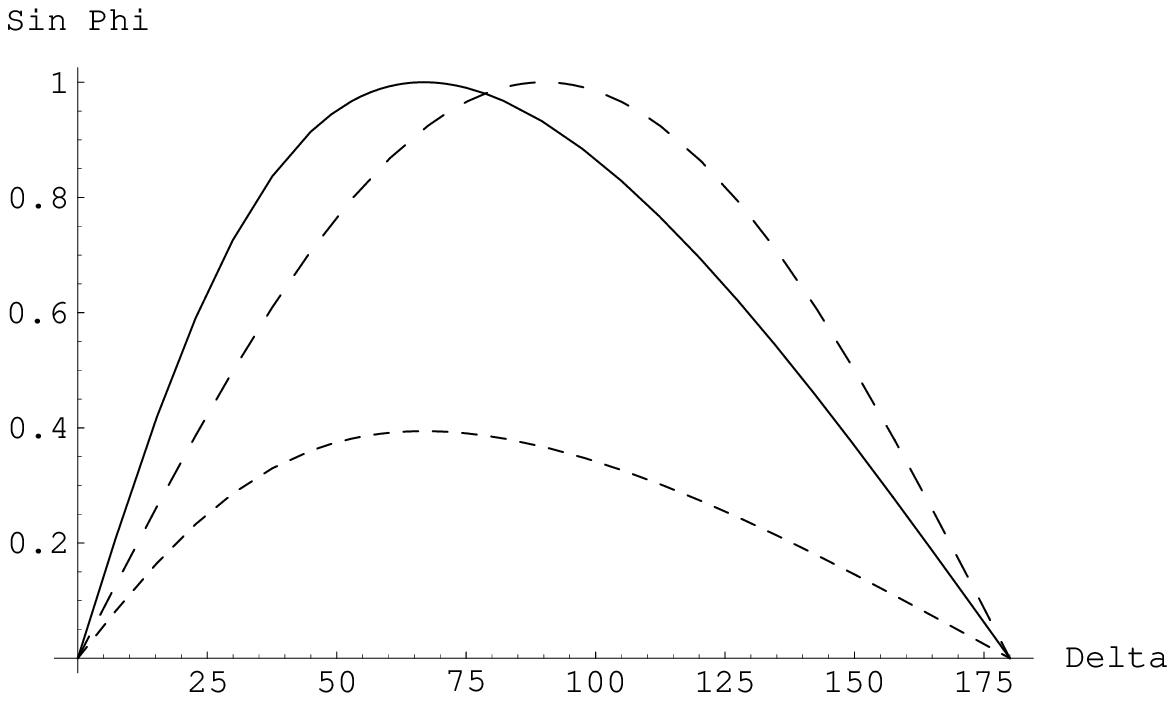}

\end{center}
\caption{
$\sin\phi_i$ ($i=1,2,3$) versus $\delta$ in $V(1,3)$.  
The curves $\sin\alpha$, $\sin\beta$ and $\sin\gamma$ are
denoted by a solid line, a dotted line  and  a dashed line,
respectively.
}
\label{v13}
\end{figure}

\begin{figure}
\begin{center}

\includegraphics[width=8.6cm]{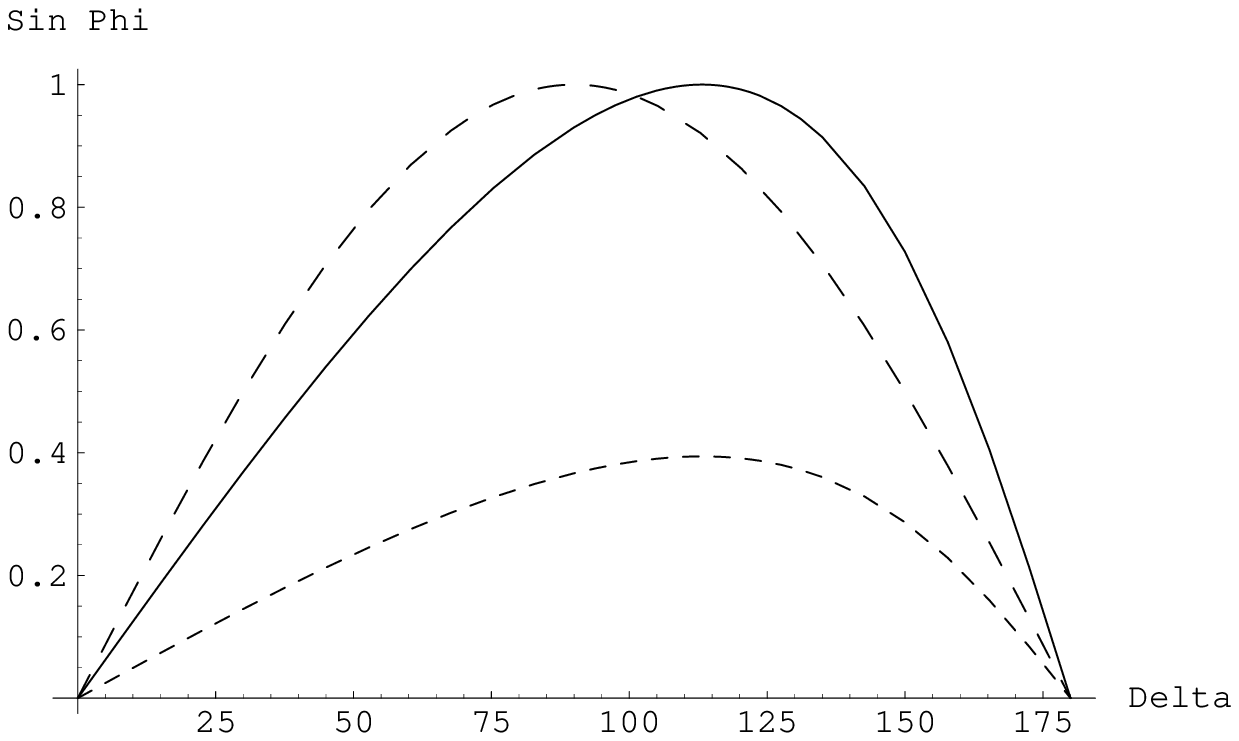}

\end{center}
\caption{
$\sin\phi_i$ ($i=1,2,3$) versus $\delta$ in $V(2,3)$.  
The curves $\sin\alpha$, $\sin\beta$ and $\sin\gamma$ are
denoted by a solid line, a dotted line  and  a dashed line,
respectively.
}
\label{v23}
\end{figure}

\begin{figure}
\begin{center}

\includegraphics[width=8.6cm]{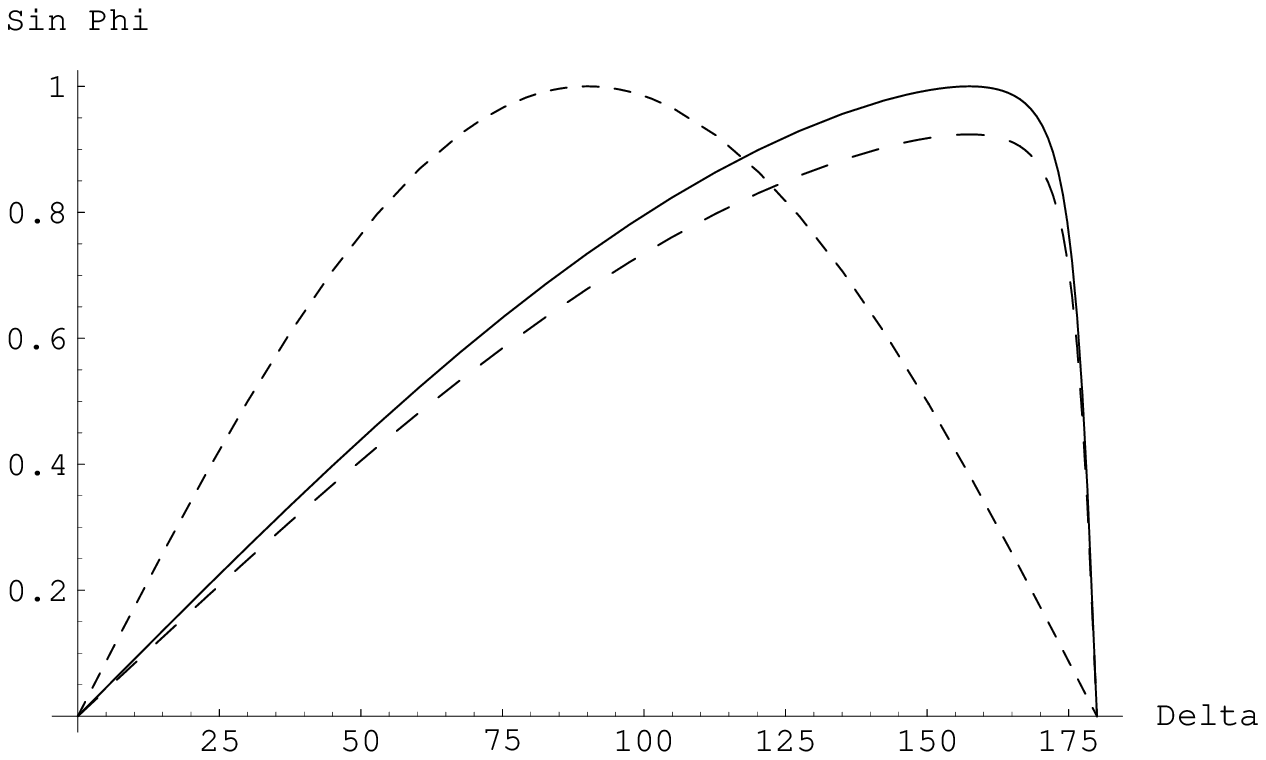}

\end{center}
\caption{
$\sin\phi_i$ ($i=1,2,3$) versus $\delta$ in $V(2,1)$.  
The curves $\sin\alpha$, $\sin\beta$ and $\sin\gamma$ are
denoted by a solid line, a dotted line  and  a dashed line,
respectively.
}
\label{v21}
\end{figure}

\begin{figure}
\begin{center}

\includegraphics[width=8.6cm]{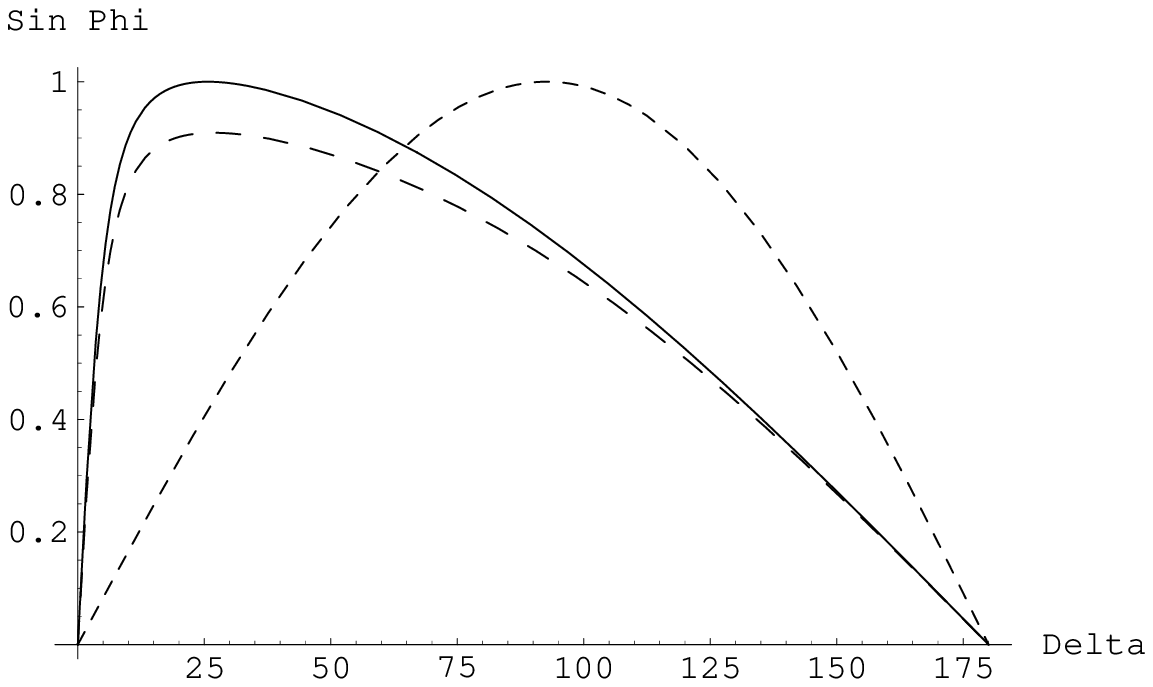}

\end{center}
\caption{
$\sin\phi_i$ ($i=1,2,3$) versus $\delta$ in $V(3,1)$.  
The curves $\sin\alpha$, $\sin\beta$ and $\sin\gamma$ are
denoted by a solid line, a dotted line  and  a dashed line,
respectively.
}
\label{v31}
\end{figure}

\begin{figure}
\begin{center}

\includegraphics[width=8.6cm]{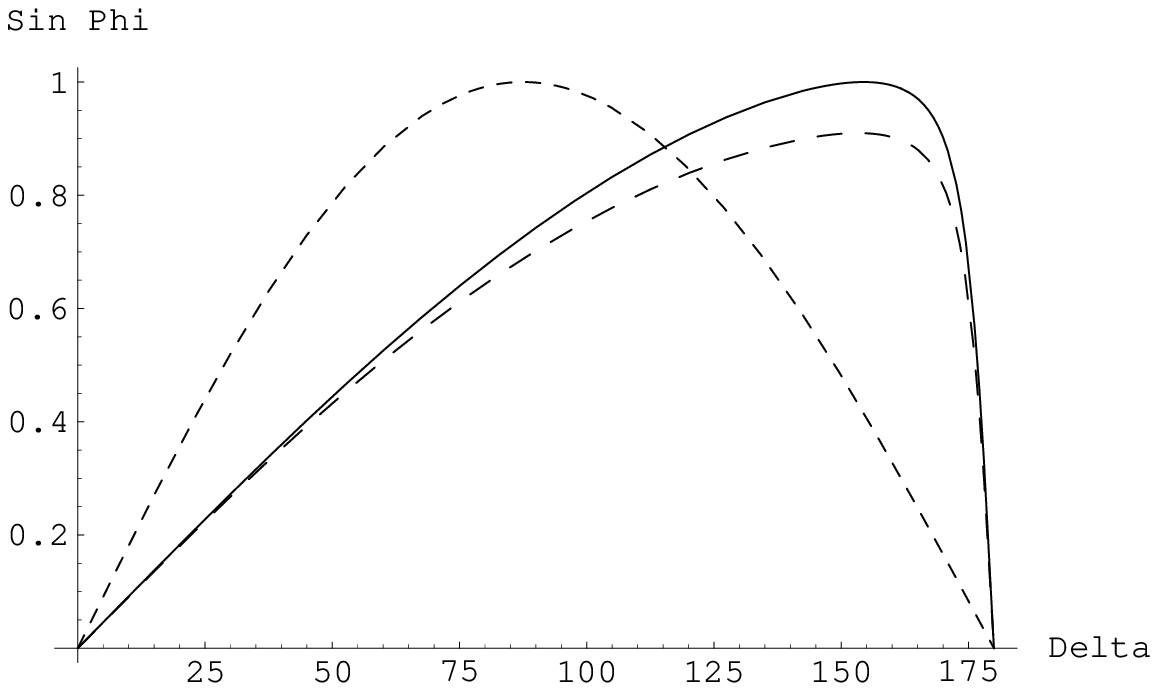}

\end{center}
\caption{
$\sin\phi_i$ ($i=1,2,3$) versus $\delta$ in $V(3,2)$.  
The curves $\sin\alpha$, $\sin\beta$ and $\sin\gamma$ are
denoted by a solid line, a dotted line  and  a dashed line,
respectively.
}
\label{v32}
\end{figure}

\end{document}